\documentclass[prb,twocolumn,showpacs,preprintnumbers,amsmath,amssymb,superscriptaddress]{revtex4-1}

\usepackage{graphicx}
\usepackage{dcolumn}
\usepackage{bm}

\begin{document}

\title{Analysis of magnetic neutron-scattering data of two-phase ferromagnets}

\author{Dirk~Honecker}\email[Corresponding author. Electronic address: ]{dirk.honecker@uni.lu}
\affiliation{Physics and Materials Science Research Unit, University of Luxembourg, 162A~Avenue de la Fa\"iencerie, L-1511 Luxembourg, Grand Duchy of Luxembourg}
\author{Charles~D.~Dewhurst}
\affiliation{Institut Laue-Langevin, 6~Rue Jules Horowitz, B.P.~156, F-38042~Grenoble Cedex~9, France}
\author{Kiyonori~Suzuki}
\affiliation{Department of Materials Engineering, Monash University, Clayton, Victoria~3800, Australia}
\author{Sergey Erokhin}
\address{INNOVENT Technology Development, Pr\"ussingstra{\ss}e 27B, D-07745 Jena, Germany}
\author{Andreas~Michels}
\affiliation{Physics and Materials Science Research Unit, University of Luxembourg, 162A~Avenue de la Fa\"iencerie, L-1511 Luxembourg, Grand Duchy of Luxembourg}

\date{\today}

\begin{abstract}
We have analyzed magnetic-field-dependent small-angle neutron scattering (SANS) data of soft magnetic two-phase nanocomposite ferromagnets in terms of a recent micromagnetic theory for the magnetic SANS cross section [D.~Honecker and A.~Michels, Phys.~Rev.~B $\mathbf{87}$, 224426 (2013)]. The approach yields a value for the average exchange-stiffness constant and provides the Fourier coefficients of the magnetic anisotropy field and magnetostatic field, which is related to jumps of the magnetization at internal interfaces.
\end{abstract}


\maketitle\

Progress in the field of nanomagnetism \cite{bader06} relies on the continuous development and improvement of observational (microscopy and scattering) techniques. For instance, advances in spin-polarized scanning tunneling microscopy, electron microscopy and holography, Kerr microscopy, and synchrotron-based x-ray techniques such as x-ray magnetic circular dichroism allows one to resolve ever finer details of the magnetic microstructure of materials, with a spatial resolution that ranges from macroscopic dimensions down to the atomic scale (see, e.g., Ref.~\onlinecite{obstechniques} and references therein).

The technique of neutron scattering is of particular importance for magnetism investigations, since it provides access to the structure and dynamics of magnetic materials on a wide range of length and time scales. \cite{tapan2006} Moreover, in contrast to electrons or light, neutrons are able to penetrate deeply into matter and, thus, enable the study of bulk properties.

Magnetic small-angle neutron scattering (SANS) measures the diffuse scattering along the forward direction which arises from nanoscale variations in the magnitude and orientation of the magnetization vector field $\mathbf{M}(\mathbf{r})$. \cite{fitz04,kohlbrecher05,michels08rop,albi2010} The measurable quantity in a magnetic SANS experiment---the (energy-integrated) macroscopic differential scattering cross section $d \Sigma / d \Omega$---depends on the Fourier coefficients of $\mathbf{M}(\mathbf{r})$. These Fourier coefficients $\widetilde{\mathbf{M}}(\mathbf{q})$ depend in a complicated manner on the magnetic interactions, the underlying microstructure (e.g., particle-size distribution and crystallographic texture), and on the applied magnetic field. The continuum theory of micromagnetics \cite{brown,aharonibook,kronfahn03} provides the proper framework for computing $d \Sigma / d \Omega$. \cite{erokhin2012prb,michels2012prb1,michels2013jmmm}

In a recent paper \cite{michels2013} we have derived closed-form expressions for the micromagnetic SANS cross section of two-phase particle-matrix-type bulk ferromagnets. Prototypical examples for this class of materials are hard and soft magnetic nanocomposite magnets, which consist of a dispersion of crystalline nanoparticles in a (crystalline or amorphous) magnetic matrix. Due to their technological relevance, e.g., as integral components in electronics devices or motors, these materials are the subject of an intense worldwide research effort. \cite{suzuki06,gutfleisch2011} From the micromagnetic point of view, due to the change of the materials parameters (exchange and anisotropy constants, saturation magnetization), the internal interfaces cause a significant perturbation of the magnetization distribution. In fact, previous magnetization and electron-holography studies \cite{he94,varga2000,li02,gao03} have discussed the effect of magnetostatic interactions in such samples, and SANS experiments \cite{michels05epl,michels06prb,elmas09} have indicated that jumps in the value of the saturation magnetization at the particle-matrix interface represent a dominating source of spin disorder.

It is the aim of this communication to test the previously published theory for the magnetic SANS cross section of two-phase nanocomposites (Ref.~\onlinecite{michels2013}) against experimental data and to determine quantitatively the magnetic-interaction parameters; in particular, the exchange constant and the strength and spatial structure of the magnetic anisotropy and magnetostatic field. For this purpose, we have analyzed existing magnetic-field-dependent neutron data of soft magnetic nanocomposites from the Nanoperm family of alloys. \cite{michels05epl,michels06prb,michels2012prb2} The microstructure of these materials consists of a dispersion of bcc iron nanoparticles in an amorphous magnetic matrix. \cite{suzuki06} The particular alloys under study have a nominal composition of $\rm{Fe}_{89}\rm{Zr}_7\rm{B}_3\rm{Cu}$ (particle size: $12 \pm 2 \, \mathrm{nm}$; crystalline volume fraction: $\sim 40 \, \%$; saturation magnetization: $1.26 \, \mathrm{T}$) and $(\rm{Fe}_{0.985}\rm{Co}_{0.015})_{90}\rm{Zr}_7\rm{B}_3$ (particle size: $15 \pm 2 \, \mathrm{nm}$; crystalline volume fraction: $\sim 65 \, \%$; saturation magnetization: $1.64 \, \mathrm{T}$); the addition of a small amount of Co results in a vanishing magnetostriction. \cite{suzuki1994} For more details on sample synthesis, characterization, and on the SANS experiments, we refer to Refs.~\onlinecite{suzuki1994,suzuki06,michels05epl,michels06prb,michels2012prb2}.

As shown in Ref.~\onlinecite{michels2013}, near magnetic saturation and for the scattering geometry where the applied magnetic field $\mathbf{H}_0 \parallel \mathbf{e}_z$ is perpendicular to the wave vector of the incoming neutron beam, the elastic SANS cross section $d \Sigma / d \Omega$ of a two-phase particle-matrix-type ferromagnet can be written as
\begin{equation}
\label{sigmasansperp2d}
\frac{d \Sigma}{d \Omega}(\mathbf{q}) = \frac{d \Sigma_{\mathrm{res}}}{d \Omega}(\mathbf{q}) + \frac{d \Sigma_M}{d \Omega}(\mathbf{q}) ,
\end{equation}
where
\begin{equation}
\label{sigmaresperp}
\frac{d \Sigma_{\mathrm{res}}}{d \Omega}(\mathbf{q}) = \frac{8 \pi^3}{V} \, b_H^2 \left( \frac{|\widetilde{N}|^2}{b_H^2} + |\widetilde{M}_z|^2 \sin^2\theta \right)
\end{equation}
represents the (nuclear and magnetic) residual SANS cross section, which is measured at complete magnetic saturation, and
\begin{equation}
\label{sigmasmperp}
\frac{d \Sigma_M}{d \Omega}(\mathbf{q}) = S_H(\mathbf{q}) \, R_H(q, \theta, H_i) + S_M(\mathbf{q}) \, R_M(q, \theta, H_i)
\end{equation}
is the spin-misalignment SANS cross section. In these expressions, $\mathbf{q}$ is the momentum-transfer vector, $V$ is the sample volume, $b_H = 2.9 \times 10^{8} \, \mathrm{A}^{-1} \mathrm{m}^{-1}$, and $\widetilde{N}(\mathbf{q})$ and $\widetilde{M}_z(\mathbf{q})$ denote, respectively, the Fourier amplitudes of the nuclear scattering-length density and of the longitudinal magnetization (parallel to $\mathbf{H}_0$). The magnetic scattering due to transversal spin components, with related Fourier amplitudes $\widetilde{M}_x(\mathbf{q})$ and $\widetilde{M}_y(\mathbf{q})$, is contained in $d \Sigma_M / d \Omega$, which decomposes into a contribution $S_H \times R_H$ due to perturbing magnetic anisotropy fields and a part $S_M \times R_M$ related to magnetostatic fields. The micromagnetic SANS theory considers a uniform exchange interaction and a random distribution of magnetic easy axes, but takes explicitely into account variations in the magnitude of the magnetization (via the function $S_M$, see below).

The anisotropy-field scattering function
\begin{equation}
\label{shdef}
S_H(\mathbf{q}) = \frac{8 \pi^3}{V} \, b_H^2 \, |h|^2
\end{equation}
depends on the Fourier coefficient $h(\mathbf{q})$ of the magnetic anisotropy field, whereas the scattering function of the longitudinal magnetization
\begin{equation}
\label{smdef}
S_M(\mathbf{q}) = \frac{8 \pi^3}{V} \, b_H^2 \, |\widetilde{M}_z|^2
\end{equation}
provides information on the magnitude $\Delta M \propto \widetilde{M}_z$ of the magnetization jump at internal (particle-matrix) interfaces. The corresponding (dimensionless) micromagnetic response functions can be expressed as
\begin{equation}
\label{rhdefperp}
R_H(q, \theta, H_i) = \frac{p^2}{2} \left( 1 + \frac{\cos^2\theta}{\left( 1 + p \sin^2\theta \right)^2} \right)
\end{equation}
and
\begin{equation}
\label{rmdefperp}
R_M(q, \theta, H_i) = \frac{p^2 \, \sin^2\theta \cos^4\theta}{\left( 1 + p \sin^2\theta \right)^2} + \frac{2 p \, \sin^2\theta \cos^2\theta}{1 + p \sin^2\theta} ,
\end{equation}
where $p(q, H_{\rm i}) = M_{\rm s} / H_{\rm eff}$ and $\theta$ represents the angle between $\mathbf{H}_0$ and $\mathbf{q} \cong q (0, \sin\theta, \cos\theta)$. The effective magnetic field $H_{\rm eff}(q, H_i) = H_i \left( 1 + l_H^2 q^2 \right)$ depends on the internal magnetic field $H_i$ and on the exchange length $l_H(H_i) = \sqrt{2 A / (\mu_0 M_s H_i)}$ ($M_s$: saturation magnetization; $A$: exchange-stiffness parameter; $\mu_0 = 4\pi 10^{-7} \, \mathrm{T m/A}$). When the functions $\widetilde{N}$, $\widetilde{M}_z$ and $h$ depend only on the magnitude $q$ of the scattering vector, one can perform an azimuthal average of Eq.~(\ref{sigmasansperp2d}). The resulting expressions for the response functions read
\begin{equation}
\label{rhdefperpradav}
R_H(q, H_i) = \frac{p^2}{4} \left( 2 + \frac{1}{\sqrt{1 + p}} \right)
\end{equation}
and
\begin{equation}
\label{rmdefperpradav}
R_M(q, H_i) = \frac{\sqrt{1 + p} - 1}{2} ,
\end{equation}
so that the azimuthally-averaged total nuclear and magnetic SANS cross section can written as
\begin{equation}
\label{sigmafinal}
\frac{d \Sigma}{d \Omega}(q) = \frac{d \Sigma_{\mathrm{res}}}{d \Omega}(q) + S_H(q) \, R_H(q, H_i) + S_M(q) \, R_M(q, H_i) .
\end{equation}

\begin{figure}[tb]
\centering
\resizebox{0.80\columnwidth}{!}{\includegraphics{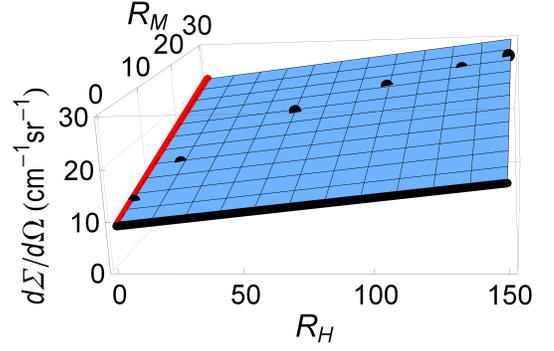}}
\caption{(Color online) Illustration of the neutron-data analysis procedure according to Eq.~(\ref{sigmafinal}). The total $d \Sigma / d \Omega$ ($\bullet$) at $q^{\star} = 0.114 \, \mathrm{nm}^{-1}$ is plotted versus the response functions $R_H$ and $R_M$ at experimental field values (in mT) of 1270, 312, 103, 61, 42, 33. The plane represents a fit to Eq.~(\ref{sigmafinal}). The intercept of the plane with the $d \Sigma / d \Omega$-axis provides the residual SANS cross section $d \Sigma_{\mathrm{res}} / d \Omega$, while $S_H$ and $S_M$ are obtained from the slopes of the plane (slopes of the thick black and red lines).}
\label{fig1}
\end{figure}

For given values of the materials parameters $A$ and $M_s$, the numerical values of both response functions are known at each value of $q$ and $H_i$. Equation~(\ref{sigmafinal}) is linear in both $R_H$ and $R_M$, with \textsl{a priori} unknown functions $d \Sigma_{\mathrm{res}} / d \Omega$, $S_H$ and $S_M$. By plotting at a particular $q = q^{\star}$ the values of $d \Sigma / d \Omega$ measured at several $H_i$ versus $R_H(q^{\star}, H_i, A)$ and $R_M(q^{\star}, H_i, A)$, one can obtain the values of $d \Sigma_{\mathrm{res}} / d \Omega$, $S_H$ and $S_M$ at $q = q^{\star}$ by (weighted) least-squares plane fits (see Fig.~\ref{fig1}). Treating the exchange-stiffness constant in the expression for $H_{\rm eff}$ as an adjustable parameter, allows one to obtain information on this quantity.

\begin{figure}[htb]
\centering
\resizebox{0.80\columnwidth}{!}{\includegraphics{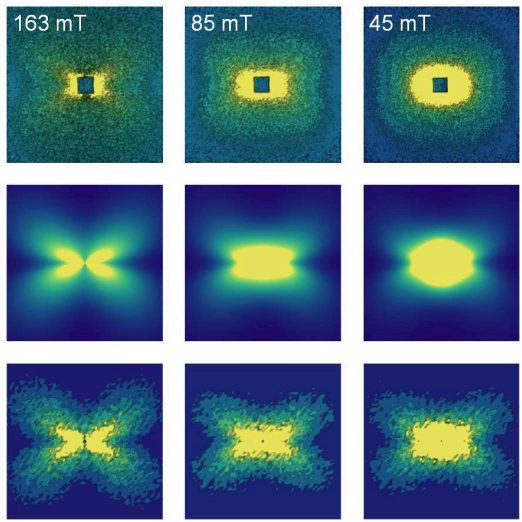}}
\caption{(Color online) Qualitative comparison between experiment, analytical theory, and numerical micromagnetic simulations. Upper row: Experimental spin-misalignment SANS cross sections $d \Sigma_M / d \Omega$ of $\rm{Fe}_{89}\rm{Zr}_7\rm{B}_3\rm{Cu}$ (Ref.~\onlinecite{michels05epl}) in the plane of the two-dimensional detector at selected applied magnetic fields (see insets). The $d \Sigma_M / d \Omega$ were obtained by subtracting the scattering at a saturating field of $1994 \, \mathrm{mT}$. $\mathbf{H}_0$ is horizontal. Middle row: Prediction by the micromagnetic theory for $d \Sigma_M / d \Omega$, Eq.~(\ref{sigmasmperp}), at the same field values as above. For both $\widetilde{M}_z(qR)$ and $h(qR)$ we have used the form factor of the sphere with a radius of $R = 6 \, \mathrm{nm}$. Furthermore, the following materials parameters were used: $A = 3.1 \, \mathrm{pJ/m}$; $\mu_0 M_s = 1.26 \, \mathrm{T}$; $\mu_0 H_p = 0.01 \, \mathrm{T}$; $\mu_0 \Delta M = 0.05 \, \mathrm{T}$. Lower row: Results of full-scale three-dimensional micromagnetic simulations for $d \Sigma_M / d \Omega$; for further details, see Refs.~\onlinecite{erokhin2012prb,michels2012prb1,michels2013jmmm}. Pixels in the corners of the images have $q \cong 0.5 \, \mathrm{nm}^{-1}$. Linear color scale is used.}
\label{fig2}
\end{figure}

Figure~\ref{fig2} provides a qualitative comparison between experiment, analytical theory, and numerical micromagnetic simulations for the spin-misalignment SANS cross section $d \Sigma_M / d \Omega$. The purpose of this figure is to demonstrate that the experimental anisotropy ($\theta$-dependence) of $d \Sigma_M / d \Omega$ (upper row in Fig.~\ref{fig2}) can be well reproduced by the theory. At the largest fields, one observes the so-called clover-leaf anisotropy with maxima in $d \Sigma_M / d \Omega$ roughly along the diagonals of the detector. Clearly, this feature is due to the term $S_M \times R_M$ in $d \Sigma_M / d \Omega$ [compare Eq.~(\ref{rmdefperp})]. Reducing the field results in the emergence of a scattering pattern that is more of a $\cos^2\theta$-type (with maxima along the horizontal direction). The observed transition in the experimental data is qualitatively reproduced by the analytical micromagnetic theory (middle row, compare also Figs.~2 and 3 in Ref.~\onlinecite{michels2013}), and by the results of full-scale three-dimensional micromagnetic simulations for $d \Sigma_M / d \Omega$ (lower row). Note that both analytical theory and micromagnetic simulations do not contain instrumental smearing effects.

\begin{figure*}[htb]
\centering
\resizebox{0.80\columnwidth}{!}{\includegraphics{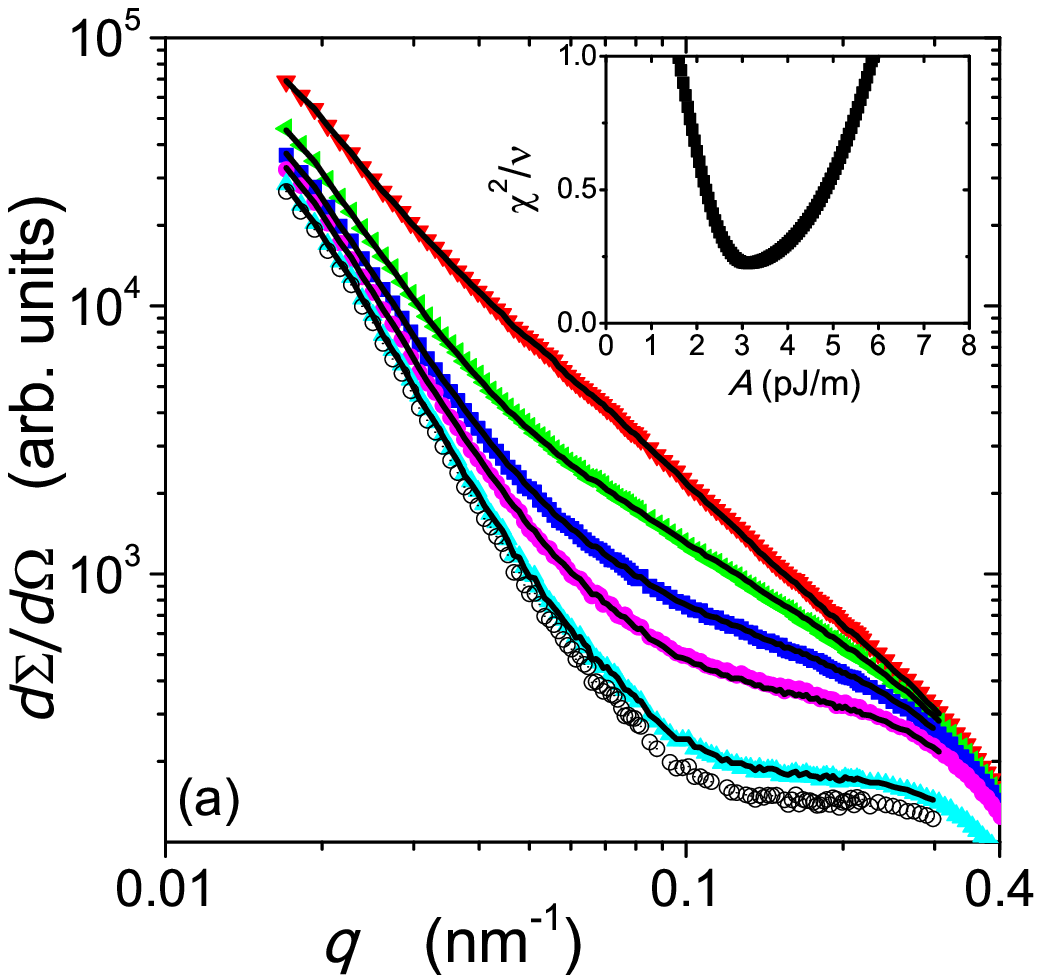}}
\resizebox{0.805\columnwidth}{!}{\includegraphics{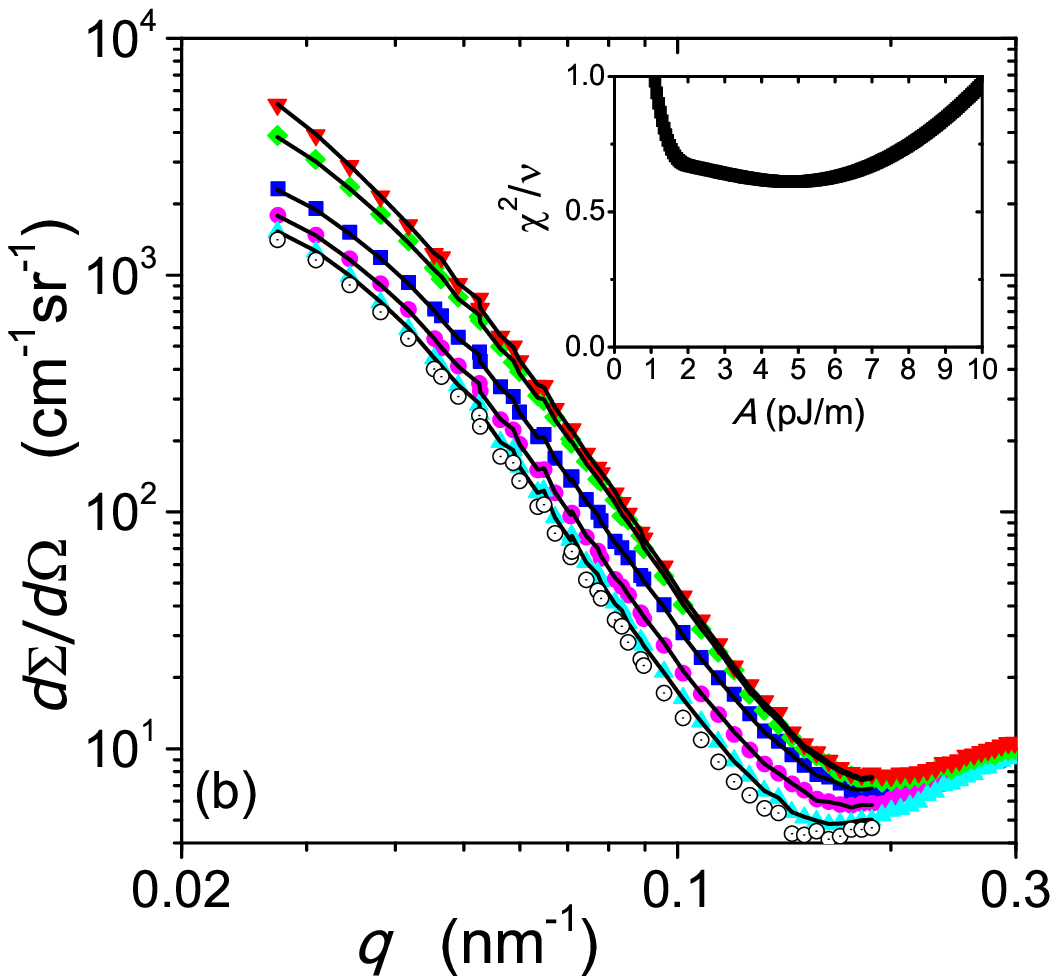}}
\caption{(Color online) Azimuthally-averaged $d \Sigma / d \Omega$ of (a) $\rm{Fe}_{89}\rm{Zr}_7\rm{B}_3\rm{Cu}$ (Ref.~\onlinecite{michels05epl}) and (b) $(\rm{Fe}_{0.985}\rm{Co}_{0.015})_{90}\rm{Zr}_7\rm{B}_3$ (Ref.~\onlinecite{michels2012prb2}) at selected applied magnetic fields (log-log scale). Field values (in mT) from bottom to top: (a) 1994, 321, 163, 85, 45; (b) 1270, 312, 103, 61, 33. Solid lines in (a) and (b): Fit to the micromagnetic theory, Eq.~(\ref{sigmafinal}). ($\circ$) Residual scattering cross sections $d \Sigma_{\mathrm{res}} / d \Omega$. The insets depict the respective (reduced) weighted mean-square deviation between experiment and fit, $\chi^2/\nu$, as a function of the exchange-stiffness constant $A$.}
\label{fig3}
\end{figure*}

The azimuthally-averaged field-dependent SANS cross sections of both Nanoperm samples along with the fits to the micromagnetic theory [Eq.~(\ref{sigmafinal}), solid lines] are displayed in Figs.~\ref{fig3}(a) and \ref{fig3}(b).
It is seen that for both samples the entire ($q, H$) dependence of $d \Sigma / d \Omega$ can be excellently described by the micromagnetic prediction. As expected, both residual SANS cross sections $d \Sigma_{\mathrm{res}} / d \Omega$ ($\circ$) are smaller than the respective total $d \Sigma / d \Omega$ at the highest field, supporting the notion of dominant spin-misalignment scattering in these type of materials. From the fit of the entire $(q, H)$ data set to Eq.~(\ref{sigmafinal}) one obtains values for the volume-averaged exchange-stiffness constants [compare insets in Figs.~\ref{fig3}(a) and \ref{fig3}(b)]. We obtain $A = 3.1 \pm 0.1 \, \mathrm{pJ/m}$ for the Co-free alloy and $A = 4.7 \pm 0.9 \, \mathrm{pJ/m}$ for the zero-magnetostriction Nanoperm sample.

Since jumps in $A$ have not been taken into account in our micromagnetic SANS theory, the determined $A$-values represent mean values, averaged over crystalline and amorphous regions within the sample. The thickness $\delta$ of the intergranular amorphous layer between the iron nanoparticles can be roughly estimated by \cite{herzer97} $\delta = D (x^{-1/3}_C- 1)$, where $D$ is the average particle size and $x_C$ denotes the crystalline volume fraction. For $\rm{Fe}_{89}\rm{Zr}_7\rm{B}_3\rm{Cu}$ with $D = 12 \, \mathrm{nm}$ and $x_C = 40 \, \%$ we obtain $\delta \cong 4 \, \mathrm{nm}$, whereas $\delta \cong 2 \, \mathrm{nm}$ for $(\rm{Fe}_{0.985}\rm{Co}_{0.015})_{90}\rm{Zr}_7\rm{B}_3$ with $D = 15 \, \mathrm{nm}$ and $x_C = 65 \, \%$. Since one may expect that the effective exchange stiffness is governed by the weakest link in the bcc-amorphous-bcc coupling chain, \cite{suzuki98a,suzuki06} the above determined experimental values for $A$ reflect qualitatively the trend in $\delta$ (and hence in $x_C$) between the two samples.

The experimental $A$-values seem to be in agreement with the following expression for the effective exchange-stiffness constant $A$ of two-phase magnetic nanostructures, \cite{suzuki06}
\begin{equation}
\label{Aformular}
\frac{D + \delta}{\sqrt{A}} = \frac{D}{\sqrt{A_{\mathrm{cr}}}} + \frac{\delta}{\sqrt{A_{\mathrm{am}}}},
\end{equation}
where $A_{\mathrm{cr}}$ and $A_{\mathrm{am}}$ denote, respectively, the \textsl{local} exchange constants of the crystalline iron and amorphous matrix phase. Equation~(\ref{Aformular}) has been derived by considering the behaviour of the tilting angle between exchange-coupled local magnetizations. \cite{suzuki98a} $A_{\mathrm{am}}$ can be roughly estimated by means of $A_{\mathrm{am}} = A_{\mathrm{cr}} \, \left( \frac{M_s^{\mathrm{am}}}{M_s^{\mathrm{cr}}} \right)^2 \, \frac{T_c^{\mathrm{am}}}{T_c^{\mathrm{cr}}}$, where $A_{\mathrm{cr}} = 10 \, \mathrm{pJ/m}$, $\mu_0 M_s^{\mathrm{cr}} = 2.15 \, \mathrm{T}$, $T_c^{\mathrm{cr}} = 1043 \, \mathrm{K}$ (Ref.~\onlinecite{skomski2003}), and $T_c^{\mathrm{am}} \cong 350 \, \mathrm{K}$ (Ref.~\onlinecite{suzuki98a}); $M_s^{\mathrm{am}}$ is found by using the measured $M_s$-value of the compound and the crystalline volume fraction, according to $M_s^{\mathrm{am}} = (M_s - x_C M_s^{\mathrm{cr}})/(1 - x_C)$. By inserting these estimates in Eq.~(\ref{Aformular}), we finally obtain effective values of $A = 2.1 \, \mathrm{pJ/m}$ ($\rm{Fe}_{89}\rm{Zr}_7\rm{B}_3\rm{Cu}$) and $A = 4.0 \, \mathrm{pJ/m}$ [$(\rm{Fe}_{0.985}\rm{Co}_{0.015})_{90}\rm{Zr}_7\rm{B}_3$], which agree reasonably with the experimental data.

\begin{figure*}[htb]
\centering
\resizebox{0.7125\columnwidth}{!}{\includegraphics{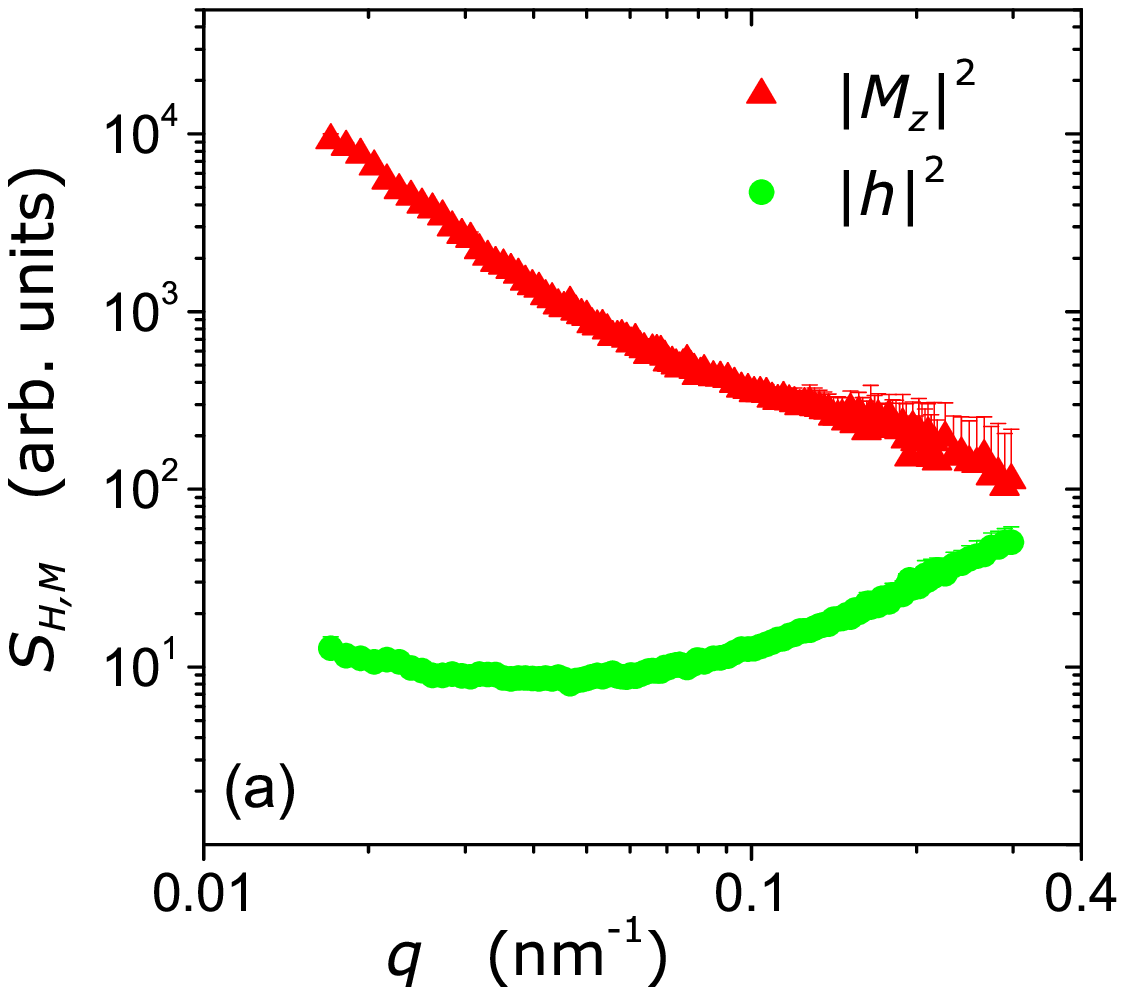}}
\resizebox{0.68\columnwidth}{!}{\includegraphics{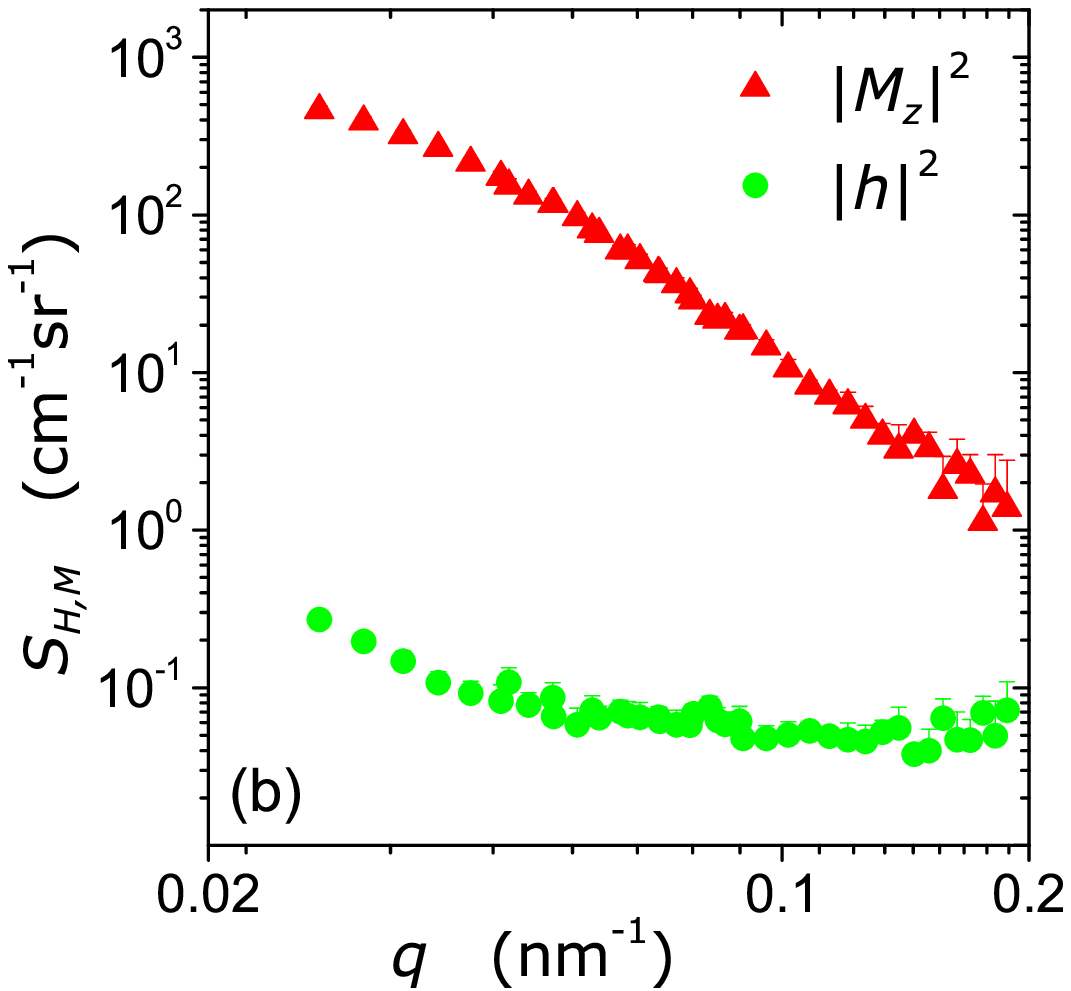}}
\caption{(Color online) Best-fit results for the scattering function of the anisotropy field $S_H = 8 \pi^3 V^{-1} b_H^2 |h(q)|^2$ and for the scattering function of the longitudinal magnetization $S_M = 8 \pi^3 V^{-1} b_H^2 |\widetilde{M}_z(q)|^2$ of (a) $\rm{Fe}_{89}\rm{Zr}_7\rm{B}_3\rm{Cu}$ and (b) $(\rm{Fe}_{0.985}\rm{Co}_{0.015})_{90}\rm{Zr}_7\rm{B}_3$ (log-log scale).}
\label{fig4}
\end{figure*}

In addition to the exchange-stiffness constant, analysis of field-dependent SANS data in terms of Eq.~(\ref{sigmafinal}) provides the magnitude squares of the Fourier coefficients of the magnetic anisotropy field $S_H \propto |h(q)|^2$ and of the longitudinal magnetization $S_M \propto |\widetilde{M}_z(q)|^2 \propto (\Delta M)^2$ (see Fig.~\ref{fig4}).
It is immediately seen in Fig.~\ref{fig4} that over most of the displayed $q$-range $|\widetilde{M}_z|^2$ is orders of magnitude larger than $|h|^2$, suggesting that jumps $\Delta M$ in the magnetization at internal interfaces is the dominating source of spin disorder in these alloys. For $\rm{Fe}_{89}\rm{Zr}_7\rm{B}_3\rm{Cu}$ at the largest $q$, the Fourier coefficient $|h|^2$ becomes comparable to $|\widetilde{M}_z|^2$ [Fig.~\ref{fig4}~(a)]. This explains the existence of the $\cos^2\theta$-type anisotropy in $d \Sigma_M / d \Omega$ at the smallest fields (compare Fig.~\ref{fig2}).

Numerical integration of $S_H(q)$ and $S_M(q)$ over the whole $\mathbf{q}$-space, i.e.,
\begin{equation}
\label{Hpvoldef}
\frac{1}{2\pi^2 b_H^2} \int_0^{\infty} S_{H,M} \, q^2 \, dq ,
\end{equation}
yields, respectively, the mean-square anisotropy field $\langle |\mathbf{H}_p|^2 \rangle$ and the mean-square longitudinal magnetization fluctuation $\langle M_z^2 \rangle$ (e.g., Ref.~\onlinecite{jprb2001}). However, since experimental data for $S_H$ and $S_M$ are only available within a finite range of momentum transfers (between $q_{\mathrm{min}}$ and $q_{\mathrm{max}}$) and since both integrands $S_H q^2$ and $S_M q^2$ do not show signs of convergence, one can only obtain rough lower bounds for these quantities: For the $(\rm{Fe}_{0.985}\rm{Co}_{0.015})_{90}\rm{Zr}_7\rm{B}_3$ sample (for which $d \Sigma / d \Omega$ is available in absolute units), we obtain $\mu_0 \langle |\mathbf{H}_p|^2 \rangle^{1/2} \cong 10 \, \mathrm{mT}$ and $\mu_0 \langle M_z^2 \rangle^{1/2} \cong 50 \, \mathrm{mT}$. This finding qualitatively supports the notion of dominant spin-misalignment scattering due to magnetostatic fluctuations.

Finally, we note that knowledge of $S_M \propto |\widetilde{M}_z|^2$ and of the residual SANS cross section $d \Sigma_{\mathrm{res}} / d \Omega$ [Eq.~(\ref{sigmaresperp})] allows one to obtain the nuclear scattering $|\widetilde{N}|^2$ (data not shown), without using sector-averaging procedures (in unpolarized scattering) or polarization analysis. \cite{michels2010epjb}

To summarize, we have analyzed magnetic-field-dependent SANS data of iron-based soft magnetic nanocomposites in terms of a recent micromagnetic theory for the magnetic SANS cross section. The approach provides quantitative results for the mean exchange-stiffness constant as well as for the Fourier coefficients of the magnetic anisotropy field and the longitudinal magnetization. The observed angular anisotropy of the SANS pattern, in particular, the clover-leaf anisotropy, can be well reproduced by the theory. For the two Nanoperm alloys under study, we find evidence that the magnetic microstructure close to saturation is dominated by jumps in the magnetization at internal interfaces. A lower bound for the root-mean-square longitudinal magnetization fluctuation of $\sim 50 \, \mathrm{mT}$ could be estimated, as compared to a mean magnetic anisotropy field of strength $\sim 10 \, \mathrm{mT}$.

We thank Dmitry Berkov for support regarding the micromagnetic simulations. This study was financially supported by the Institut Laue-Langevin, the Deutsche Forschungsgemeinschaft (Project No.~MI~738/6-1), and by the National Research Fund of Luxembourg in the framework of ATTRACT Project No.~FNR/A09/01.

\bibliographystyle{apsrev}

\end{document}